\immediate\write18{makeindex \jobname.nlo -s nomencl.ist -o \jobname.nls}
\documentclass[12pt]{iopart}

\usepackage{soul}  % For highlighting text
\usepackage{color} % For colored text
\usepackage{cite}  % For citations

\soulregister{\cite}{1}

\usepackage{mathptmx}
\usepackage{hyperref}
\usepackage{graphicx}% Include figure files 
\usepackage{bm}% bold math 
\usepackage{amssymb}% math symbol
\usepackage{cleveref}
\usepackage{etoolbox}
\usepackage{nomencl}
\usepackage{fancyhdr}
\makenomenclature
\fancypagestyle{nomenclature}{%
	\fancyhf{} % Clear all header and footer fields
	\fancyhead[R]{\thepage} % Keep page number centered at the footer
}

\renewcommand\nomgroup[1]{%
	\item[\bfseries
	\ifstrequal{#1}{P}{Subscript}{%
	}
	]}
\crefname{equation}{}{}
\newcommand{\ud}{\ensuremath{\mathrm{d}}}

\newcommand{\bmi}{\ensuremath{\boldsymbol}}

\newcommand{\bnabla}{\ensuremath{\boldsymbol\nabla}}
\newcommand{\bcdot}{\ensuremath{\boldsymbol\cdot}}

\newcommand{\p}{\ensuremath{\partial}}
\newcommand{\beq}{\begin{equation}}
\newcommand{\eeq}{\end{equation}}
\newcommand{\eps}{\ensuremath{\epsilon}}

\newcommand{\uD}{\ensuremath{\mathrm{D}}}
\patchcmd{\numparts}{\addtocounter{equation}{1}}{\refstepcounter{equation}}{}{}
\begin{document}
	
	\title[]{Thermal-phototactic bioconvection in a forward scattering algal suspension}
	
	\author{S. K. Rajput$^{*}$, M. K. Panda, A. Rathi}
	
	\address{Department of Mathematics, PDPM Indian Institute of Information Technology, Design and Manufacturing, Jabalpur $482005$, India}
	\ead{shubh.iiitj@gmail.com}
	\vspace{10pt}
	%		\begin{indented}
	%			\item[]Jan 2024
	%		\end{indented}
	
	\begin{abstract}
		Bioconvection induced by phototaxis and thermal gradients in an anisotropic (forward) scattering algal suspension is investigated in this article. The suspension is illuminated by collimated irradiation from above and heated either from top or bottom. The linear theory is deployed on the steady state of the proposed bioconvective system and resulting eigen value problem is solved using fourth-order accurate finite-difference scheme based on Newton-Raphson-Kantorovich iteration. The results indicate that the forward scattering and heating from above (or cooling from below) in an algal suspension enhance bioconvective stability. On the other hand, heating from below enhance bioconvective instability for a fixed forward scattering coefficient.  
	\end{abstract}	
	\vspace{2pc}
	\maketitle
	\section{Introduction}
	Bioconvection refers to the spontaneous formation of convective patterns in suspensions of swimming, non-neutrally buoyant microorganisms such as algae, protozoa, and bacteria~\cite{19platt1961,20pedley1992,21hill2005,22bees2020,23javadi2020,44rajput2024mathematical}. These microbes typically exhibit upward swimming tendencies and are denser than the surrounding fluid, leading to gravitational instabilities that drive the bioconvective process. The buoyant forces generated by their collective motility serve as the primary energy source sustaining these convective motions. In contrast, non-motile organisms generally do not exhibit such pattern formation. However, exceptions have been noted, where classical assumptions of up-swimming and excess density are not strictly required for bioconvection to occur~\cite{20pedley1992,28hill1989}. Although bioconvection is extensively studied in laboratory settings~\cite{40Sommer2017}, in situ evidence, such as micropatches of zooplankton, also supports its occurrence in natural aquatic ecosystems~\cite{36kils}. The phenomenon is often mediated by \textit{taxis}, a behavioral mechanism that enables microbes to respond to external stimuli. Microorganisms may orient themselves based on biochemical and physical cues such as light (phototaxis), chemicals (chemotaxis), gravity (gravitaxis), and flow-induced torques (gyrotaxis). This study focuses on phototactic bioconvection, where light-induced motility governs pattern formation. 
	
	In natural aquatic environments, the behavior of phototactic microorganisms, particularly in forward-scattering algal suspensions, is significantly influenced by the availability and direction of light. Oblique or vertical collimated solar flux, along with diffuse radiation, can result in markedly different bioconvective dynamics~\cite{1wager1911,3kessler1985,4williams2011,25kessler1997,26kage2013,24kessler1986,27mendelson1998,2kitsunezaki2007,5kessler1989}. In some cases, intense illumination may suppress or disrupt bioconvective patterns altogether. Several mechanisms can explain how light modulates these dynamics. First, photosynthesis drives the energy acquisition of algae, and forward scattering under oblique light may alter algal orientation as they seek optimal illumination~\cite{6hader1987}. Second, collective behaviors such as self-shading and light scattering introduce heterogeneity into the aquatic medium, affecting phototactic efficiency and inter-microbial competition~\cite{8panda2020}. The morphology of algae further determines whether the scattering is isotropic or anisotropic. Most often, light scattering is forward-biased due to size-dependent optical properties~\cite{34privoznik1978absorption,35pilon2008}.
		
	\begin{figure}[!h]
		\begin{center}
			\includegraphics[scale=0.75]{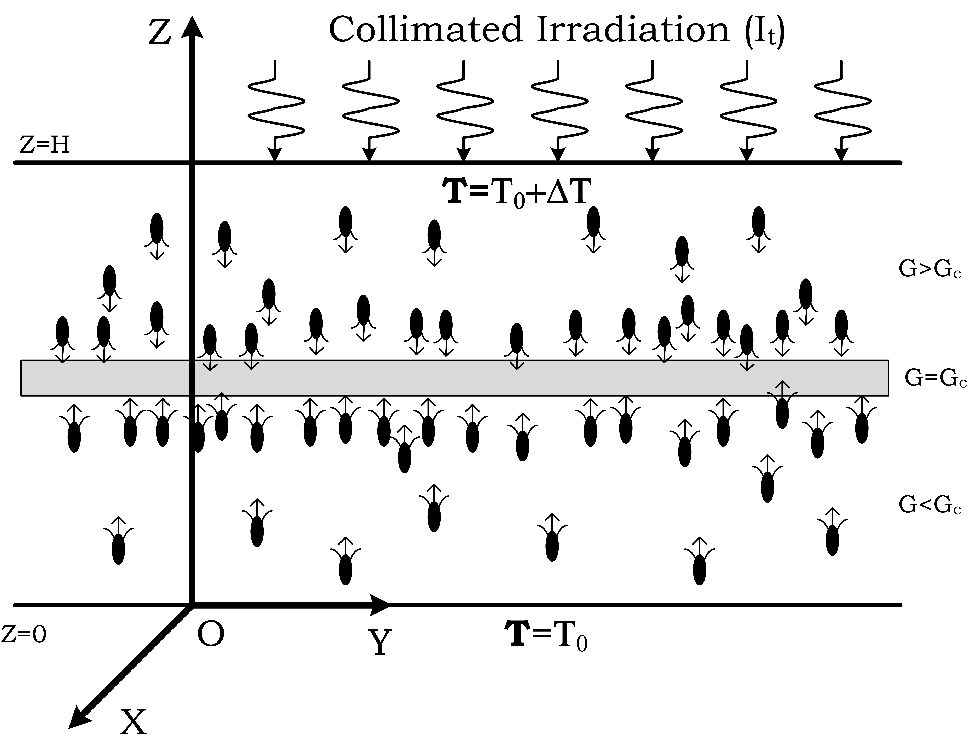}
		\end{center}
		\caption{Diagram of concentrated horizontal layer at $G=G_c.$}
		\label{fig1}
	\end{figure}  	
	
	In this work, we investigate the linear stability of a finite-depth suspension of forward-scattering phototactic microbes subject to the collimated irradiation. We consider perturbations around a steady state characterized by a balance between phototactic alignment and random reorientation of microorganisms in a quiescent fluid. This steady state features a concentrated horizontal layer (at which $G = G_c$) dividing the domain into a lower phototactically unstable region ($G < G_c$) and an upper stable region ($G > G_c$), as illustrated in Fig.\ref{fig1}. At critical parameter values (e.g., Rayleigh number), convective motions from the unstable region can penetrate the upper region, forming a classic example of penetrative convection\cite{9straughan1993,10ghorai2005,11panda2016}.
	
	Theoretical investigations into bioconvection began with the linear stability analysis of phototactic suspensions with constant thermal properties, notably by Vincent and Hill~\cite{12vincent1996}. Ghorai and Hill~\cite{10ghorai2005} further simulated these models in two-dimensional geometries, although they neglected scattering effects. Later, Ghorai et al.\cite{7ghorai2010} explored isotropic scattering and found that scattering induces bimodal steady states. Ghorai and Panda\cite{13ghorai2013} extended this to forward-scattering suspensions, highlighting the emergence of complex pattern-forming instabilities. Subsequent studies examined related problems under various illumination conditions and geometries~\cite{8panda2020,14panda2013,15panda2016,11panda2016,16panda2022,32panda2023,42rajput2025b,38rajput2024,41rajput2025a,43rajput2025c}.While these works largely focused on phototactic bioconvection with constant thermal conditions, temperature fluctuations are known to significantly affect microbial behavior and aquatic ecosystem dynamics. Many microorganisms, including thermophilic algae, thrive in environments with pronounced temperature gradients, such as hot springs or stratified ocean layers. Temperature can influence marine productivity, microbial orientation, and buoyancy. In summer, solar heating causes surface layers to warm and stabilize, making the upper layers both irradiated and thermally stratified. Under such conditions, it becomes essential to consider the combined effects of phototaxis and thermal gradients on bioconvection. Despite considerable progress in phototactic bioconvection theory, no study to date has addressed the combined effects of collimated illumination and surface heating in a forward-scattering algal suspension. To bridge this gap, the present study explores the onset of thermally and phototactically driven bioconvection in a forward scattering medium illuminated obliquely.
	
	The manuscript is organized as follows: First, the mathematical form of the problem is constituted. The steady state of it is determined next. The perturbed governing system derived from the basic state is solved numerically. Finally, a discussion on the outcomes of the study is addressed.
	 
	%%%%%%%%%%%%%%%%%%%%%%%%%%%%%%%%%%%%%%%%%%%%%%%%%%%%%	
	\section{Mathematical formulation}\label{sec2}  
	
\begin{figure}[!htbp]
		\begin{center}
			\includegraphics[scale=1.]{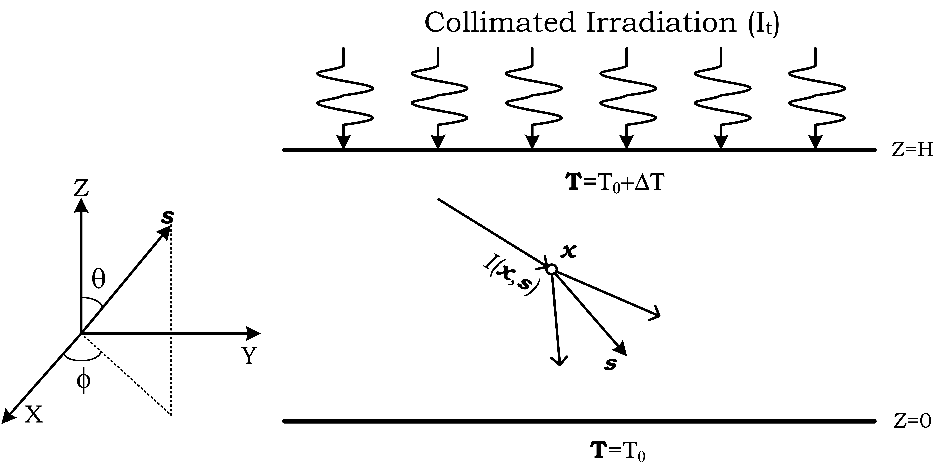}
		\end{center}
		\caption{A schematic configuration of the problem.}
		\label{fig2}
	\end{figure}

	The bioconvective system under study is shown in Fig.~\ref{fig2}. The proposed study focuses on the thermo-bioconvective instability in a forward-scattering suspension of microbes of finite thickness $H$ and infinite width. It is assumed here that the upper ($z=H$) and lower ($z=0$) boundaries do not reflect the light incident upon them. The suspension is subjected to a collimated irradiation and suspension receives heat via heat illuminating source either from the top or bottom, as illustrated in Fig.~\ref{fig2}. The temperature variation is assumed to be sufficiently mild, ensuring that it does not have lethal effects on microorganisms. Additionally, the phototactic behavior, including the orientation and speed of cell swimming, remains unaffected by the heating process. Let $L_0(\boldsymbol{x},\boldsymbol{s})$ denote the intensity of light at a given location $\boldsymbol{x}$ in the direction of $\boldsymbol{s}$.
	
	%%%%%%%%%%%%%%%%%%%%%%%%%%%%%%%%%%%%%%%%%%%%%%%%%%%%%%%%%%%%%%%%%%%%%%%%%%%%%%%%%%%%%%%%%%%%%%
	\subsection{Governing equations} \label{subsec2_1}
	
	The proposed model on phototaxis assumes that each cell within the system possesses a volume denoted by $\vartheta$ and a density $\rho+\delta\rho$, where $\rho$ represents the density of the surrounding fluid and $\delta\rho\ll\rho$ via the earlier continuum models~\cite{20pedley1992,8panda2020}. Let ${\bmi u}=(u,v,w)$ denotes the velocity and $n$ indicates concentration within the suspension. Consider the suspension to be dilute and incompressible. Hence, the mass conservation equation takes the form
	\begin{equation}
	\bnabla\bcdot{\bmi u} = 0. 
	\end{equation}
	
	Also, the momentum conservation equation is
	\begin{equation}\
	\rho\frac{\mathrm{D}{\bmi u}}{\mathrm{D} t} =  \mu{\nabla^2}{\bmi u}-\bnabla P_e  -
	n\,\delta\rho\, g \vartheta \hat {\bmi z}-\rho g(1-\beta(T-T_0))\hat{\boldsymbol{z}}, 
	\end{equation}
	where $\mu$ and $P_e$ are the suspension's viscosity (same as fluid) and excess pressure. Additionally, $\mathrm{D}/\mathrm{D}t=\p/\p t + {\bmi u}\bcdot\bnabla$ indicates the convective derivative, and $\beta$ is the coefficient of thermal expansion.
	
	Next, the equation of cell conservation is
	\begin{equation}
	\frac{\p n}{\p t} = -{\bmi \nabla}\bcdot{\bmi F}_0,
	\end{equation}
	where $\bmi F_0$ denotes the total flux of cells, determined by
	\begin{equation}
	{\bmi F}_0 = n {\bmi u}+{n} \mathbf{U}_c - \mathsf{\mathbf D} {\bmi \nabla}{n}.  
	\end{equation}
	
	Here, we consider $\mathsf{\mathbf D}=D \mathsf{\mathbf I}$ [see Panda~\cite{8panda2020}]. Therefore, 
	\beq\label{5}
	{\bmi F}_0 = n {\bmi u}+{n} \mathbf{U}_c - D {\bmi \nabla}{n}.
	\eeq
	
	Also, the temperature profile is governed by 
	\begin{equation}
	\rho_c\left(\frac{\partial T}{\partial t}+\boldsymbol{\nabla}\cdot(\boldsymbol{u}T)\right) =\alpha{\boldsymbol{\nabla}}^2 T,
	\end{equation}
	where $\rho_c$ is the volumetric heat capacity of water, and $\alpha$ is the thermal conductivity of water. In the expression of the total flux, we consider two important assumptions, which help us to discard the Fokker-Planck equation from the system [refer Panda~\cite{8panda2020} for details]
	%%%%%%%%%%%%%%%%%%%%%%%%%%%%%%%%%%%%%%%%%%%%%%%%%%%%%%%%%%%%%%%%%%%%%%%%%%%%%%%%%%%%%%%%%%%%%%	
	\subsection{The mean swimming orientation} \label{subsec2_2} 
	
	The distribution of intensity of light in a medium where microbes absorb and scatter it is expressed via a radiative transfer equation (RTE)~\cite{29modest2003,30chandrasekhar1960}
	
	%\begin{widetext}
	\beq
	{\bmi s}\bcdot\bnabla L_{0}({\bmi x},{\bmi s}) + (\kappa_0+\sigma_s)L_{0}({\bmi x},{\bmi s}) =
	\frac{\sigma_s}{4\pi}\int_0^{4\pi}L_{0}({\bmi x},{\bmi s}') g_{0}({\bmi s}',{\bmi s})\,\ud\Omega', 
	\eeq   
	%\end{widetext}
	where $\kappa_0 $ denotes absorption, and $\sigma_0$ represents the scattering coefficients. Also, $g_{0}({\bmi s}',{\bmi s})$ is a scattering phase function, which is assumed to be the linearly anisotropic~\cite{29modest2003}. Hence,
	\beq
	g_{0}({\bmi s}',{\bmi s})=1+\mathrm{A_1}({\boldsymbol{s}}'\cdot{\boldsymbol{s}}), 
	\eeq
	where $\mathrm{A_1}$ is the anisotropic scattering coefficient, which indicates backward scattering if $-1 < \mathrm{A_1} <0$, forward scattering if $0<\mathrm{A_1} \le 1$ and isotropic scattering when $\mathrm{A_1} =0$.
	
	At given any point on suspension top i.e. ${\bmi x}_H=(x,y,H)$, the light intensity is given by
	$$
	L_{0}({\bmi x}_H,{\bmi s}) = \mathrm{I_t} \,\delta({\bmi s}-{\bmi s}_0) = \mathrm{I_t}\,
	\delta(\nu_z-\nu_{z_0})\delta(\phi-\phi_0),
	$$
	where, $\mathrm{I_t}$ represents the intensity of the collimated irradiation at the top and ${\bmi s}_0=-\hat{\bmi z}$ is the direction of incidence where $\hat {\bmi z}$ represents unit vector pointed in vertical direction. 
	
	Next, we consider $\kappa_0 = \varkappa_0 n$ and $\sigma_s=\varsigma_s n$. Hence, by using scattering albedo $\omega_s=\sigma_s/(\kappa_0+\sigma_s),$ RTE converts into
	%\begin{widetext}
	\beq
	{\bmi s}\bcdot\bnabla L_{0}({\bmi x},{\bmi s}) + \beta  L_{0}({\bmi x},{\bmi s}) =\frac{\omega_s\beta }{4\pi}\int_0^{4\pi}L_{0}({\bmi x},{\bmi s}')(1+\mathrm{A_1}({\boldsymbol{s}}'\cdot{\boldsymbol{s}}))\,\ud\Omega'. 
	\eeq
	%\end{widetext}
	
	Here, $\beta=(\varkappa_0+\varsigma_s) n$ represents the extinction coefficient and $\omega_s\in[0,1]$ is the scattering albedo. Based upon the value of $\omega_s$, the suspension is characterized as purely absorbing (when $\omega_s=0$) and purely scattering (when $\omega_s=1$).
	
	The total intensity at a position ${\bmi x}$ within the suspension is expressed as
	$$
	G({\bmi x})=\int_0^{4\pi}{L_{0}({\bmi x},{\bmi s})}\,\ud\Omega,
	$$ 
	also, the radiative heat flux is defined as
	$$
	{\bmi q_{0}}({\bmi x}) = \int_0^{4\pi}{L_{0}({\bmi x},{\bmi s})}\,{\bmi s}\,\ud\Omega.
	$$  
	
	The expression for the mean swimming velocity, $\mathbf{U}_c,$ is formulated as
	$$
	\mathbf{U}_c = {U_c}\langle{\bmi p}\rangle.
	$$
	
	Here, $U_c$  signifies the speed of microbes and  their average swimming orientation $\langle{\bmi p}\rangle$ is formulated as ~\cite{8panda2020}
	\beq
	\langle{\bmi p}\rangle = -M_{0}(G) \frac{{\bmi q_{0}}}{|{\bmi q_{0}}|}, 
	\eeq
	where  $M_{0}(G)$ represents the phototaxis function,
	s.t.,
	$$
	M_{0}(G)=
	\left\{ \begin{array}{c}
	\ge 0, \qquad \mbox{when } G \le G_c, \\
	< 0, \qquad \mbox{when } G > G_c.\\
	\end{array}
	\right.     
	$$
	%%%%%%%%%%%%%%%%%%%%%%%%%%%%%%%%%%%%%%%%%%%%%%%%%%%%%%%%%%%%%%%%%%%%%%%%%%%%%%
	\subsection{Boundary conditions} \label{subsec2_3}
	
	In this study, the lower boundary is assumed to be rigid and upper boundary is stress free or rigid. Also, the cell do not move through these boundaries. Thus,
	\beq
	{\bmi u} \bcdot {\hat {\bmi z}}= {\bmi F}_0\bcdot {\hat {\bmi z}} = 0\qquad \mbox{at}\quad z=0,H,
	\eeq	
	 also, for a stress-free boundary
	 \beq
	 \frac{\partial^2}{\partial z^2}({\bmi u} \bcdot {\hat {\bmi z}})= 0\qquad \mbox{at}\quad z=0,H,
	 \eeq	 
	 and for a rigid boundary 
	 \beq
	 {\bmi u}\times {\hat {\bmi z}} = 0\qquad \mbox{at}\quad z=0.
	 \eeq
	 
	 The thermal boundary conditions are
	 \begin{equation}
	 T=T_0\qquad \mbox{at}\quad z=0.
	 \end{equation}
	 \begin{equation}
	 T=T_0+\Delta T\qquad \mbox{at}\quad z=H.
	 \end{equation}
	
	The light intensities on both boundaries are determined by
	%\begin{widetext}
	\begin{numparts}
		\begin{eqnarray}
		L_{0}(x,y,z=H,\theta,\phi) =\mathrm{I_t}
		\,\delta(\nu_z-\nu_{z_0})\delta(\phi-\phi_0),\,\quad\theta\in[\pi/2,\pi],\\
		L_{0}(x,y,z=0,\theta,\phi)=0,\, \quad \theta\in[0,\pi/2].
	\end{eqnarray}
	\end{numparts}
	%\end{widetext}
	%%%%%%%%%%%%%%%%%%%%%%%%%%%%%%%%%%%%%%%%%%%%%%%%%%%%%%%%%%%%%%%%%%%%%%%%%%%%%%
	\subsection{Scaling of the governing equations}\label{subsec2_4}  
	
	We use the same scaling parameters as used in the phototaxis model developed by Panda~\cite{8panda2020}. Hence, the governing equations become
	\begin{eqnarray}
	&& \bnabla\bcdot{\bmi u} = 0,\label{17}\\
	&& P_r^{-1}\left(\frac{\mathrm{D}{\bmi u}}{\mathrm{D} t}\right) = {\nabla^2}{\bmi u} -\bnabla P_e -
	R\,n \hat {\bmi z}-R_m\hat{\boldsymbol{z}}+R_TT\hat{\boldsymbol{z}},\\
	&& \frac{\p n}{\p t} = -{\bmi\nabla}\bcdot{\bmi F}_0, \label{19}
	\end{eqnarray}
	where ${\bmi F}_0$ is given by
	$$
	{\bmi F}_0 = n{\boldsymbol{u}}+\frac{1}{Le}nV_{c}\langle{\boldsymbol{p}}\rangle-\frac{1}{Le}{\boldsymbol{\nabla}}n, 
	$$	
	and
	\begin{equation}
	\frac{\partial T}{\partial t}+\boldsymbol{\nabla}\cdot(\boldsymbol{u}T) =\boldsymbol{\nabla}^2 T.\label{20}
	\end{equation}
	
	Here, $R=\bar n \vartheta gH^3\delta\rho/\mu \alpha_f$ is the bioconvective Rayleigh number, $R_m=gH^3\rho/\mu \alpha_f$ is the basic density Rayleigh number, $R_T=g\beta H^3\delta T/\mu \alpha_f$ is the thermal Rayleigh number, $P_r=\mu/\rho \alpha_f$ is the Schmidt number, $Le=\alpha_f/D$ is the Lewis number, $V_c=U_c H/D$ represents the non-dimensionalised swimming speed and $\alpha_f$ is the thermal density of water. 
	
	After scaling, the boundary conditions become
	\beq
	{\bmi u} \bcdot {\hat {\bmi z}}= {\bmi F}_0\bcdot {\hat {\bmi z}} = 0\qquad \mbox{at}\quad z=0,1,
	\eeq	
	for a stress-free boundary
	\beq
	\frac{\partial^2}{\partial z^2}({\bmi u} \bcdot {\hat {\bmi z}})= 0\qquad \mbox{at}\quad z=0,1,
	\eeq	 
	for a rigid boundary 
	\beq
	{\bmi u}\times {\hat {\bmi z}} = 0\qquad \mbox{at}\quad z=0.
	\eeq
	and thermal boundary conditions become
	\begin{equation}\label{24}
	T=T_0\qquad \mbox{at}\quad z=0.
	\end{equation}
	\begin{equation}\label{25}
	T=T_0+\Delta T\qquad \mbox{at}\quad z=1.
	\end{equation}
	
	Next, after scaling the RTE becomes
	%\begin{widetext}
	\begin{equation}
	{\bmi s}\bcdot\bnabla L_{0}({\bmi x},{\bmi s}) + \tau_H n L_{0}({\bmi x},{\bmi s}) =
	\frac{\omega_s\tau_H n}{4\pi}\int_0^{4\pi}L_{0}({\bmi x},{\bmi s}')(1+\mathrm{A_1}({\boldsymbol{s}}'\cdot{\boldsymbol{s}})) \,\ud\Omega',  
	\end{equation}
	%\end{widetext}
	where $\tau_H=(\varkappa_0+\varsigma_s) \bar n H$. 
	
	The RTE again converts, in the form of direction cosines $(\xi_{x},\eta_{y},\nu_{z})$ of ${\bmi s}$, as~\cite{29modest2003} 
	%\begin{widetext}
	\begin{equation}\label{27}
	\fl\xi_{x} \frac{\p L_{0}}{\p x} + \eta_{y} \frac{\p L_{0}}{\p y} +\nu_{z} \frac{\p L_{0}}{\p z} + \tau_H n
	L_{0}({\bmi x},{\bmi s})=
	\frac{\omega_s\tau_H n}{4\pi}\int_0^{4\pi}L_{0}({\bmi x},{\bmi s}')(1+\mathrm{A_1}({\boldsymbol{s}}'\cdot{\boldsymbol{s}})) \,\ud\Omega',~~~~~~~~
	\end{equation} 
	%\end{widetext}
	where 
	$$
	\xi_{x}=\sin\theta\cos\phi,\quad \eta_{y}=\sin\theta\sin\phi,\quad \nu_{z}=\cos\theta,
	$$
	with the boundary conditions
	%\begin{widetext}
	\begin{numparts}
		\begin{eqnarray}
		L_{0}(x,y,z=1,\theta,\phi) =\mathrm{I_t}
		\,\delta(\nu_z-\nu_{z_0})\delta(\phi-\phi_0),\,\quad\theta\in[\pi/2,\pi],\\
		L_{0}(x,y,z=0,\theta,\phi)=0,\, \quad \theta\in[0,\pi/2].
		\end{eqnarray}
	\end{numparts} 
	
	It is necessary to note that notations employed for the scaled and non-scaled variables are the same here. 
	
	The photoresponse curve, $M_{0}(G)$\cite{8panda2020}, can be represented by                    
	%\begin{widetext}
	\begin{equation}
	 M_{0}(G)=0.8\sin{\left(\frac{3\pi}{2}\Upsilon{\left(G\right)}\right)}-0.1\sin{\left(\frac{\pi}{2}\Upsilon{\left(G\right)}\right)},\quad
	\Upsilon{\left(G\right)}=\frac{G}{3.8} e^{\chi\left(3.8-G\right)},   
	\end{equation}
	%\end{widetext}
	where $\chi=0.252$ and the critical intensity is fixed at $1.3$.                         
	%%%%%%%%%%%%%%%%%%%%%%%%%%%%%%%%%%%%%%%%%%%%%%%%%%%%%%%%%%%%%%%%%%%%%%%%%%%%%%%%%%%%%%%%%%%%%%
	\section{The equilibrium solution}\label{sec3}
	
	To obtain the equilibrium solution of the system, we substitute
	\begin{eqnarray*}
		{\bmi u}=(0,0,0),\quad n=n_s(z), \quad \langle{\bmi
			p}\rangle=\langle{\bmi p}_{0}^{s}\rangle,\quad \textnormal{and}\quad T=T_s(z), 
	\end{eqnarray*}
	into Equations~(\ref{17})--(\ref{19}) and (\ref{27}).
	
	Therefore, the total intensity $G=G_{0}^{s}(z)$ and the radiative heat flux ${\bmi q}_{0}={\bmi q}_{0}^{s}(z)$ in the equilibrium state are determined by
	$$
	G_{0}^{s}(z)=\int_0^{4\pi}L_{0}^{s}(z,\theta)\,\ud\Omega,\quad {\bmi q}_{0}^{s}(z) =
	\int_0^{4\pi}L_{0}^{s}(z,\theta)\,{\bmi s}\,\ud\Omega.
	$$    
	
	The light intensity in the equilibrium state, $L_{0}^{s}(z,\theta)$, does not depend of $\phi$. Thus, ${\bmi q}_{0}^{s}=-q_{0}^{s}\hat{\bmi z}$, where $q_{0}^{s}=|{\bmi q}_{0}^{s}|$.
	
	Next, the light intensity, $L_0^s(z,\theta)$, at the equilibrium state is governed by
	\begin{equation}\label{30}
	\frac{\partial L_0^s}{\partial z}+\left(\frac{\tau_H n_s}{\nu_z}\right)L_0^s=\frac{\omega_s\tau_H n_s }{4\pi\nu_z}\int_0^{4\pi}L_0^s\Big(1
	+A_1({\boldsymbol{s}}'\cdot{\boldsymbol{s}})\Big)\ud\Omega'.
	\end{equation}
	
	Due to the symmetry of light intensity with respect to the azimuthal angle, the following relation holds
	\begin{equation*}
	{\boldsymbol{s}}'\cdot{\boldsymbol{s}}=\cos\theta\cos\theta',
	\end{equation*}
	where $\theta$ and $\theta'$ represent the zenith angles, defining the direction ${\boldsymbol{s}}$ and ${\boldsymbol{s}}'$ respectively. Substituting this relation into Equations~(\ref{30}) and simplifying the expression, we obtain
	\begin{equation}
	\frac{\partial L_0^s}{\partial z}+\left(\frac{\tau_H n_s}{\nu_z}\right)L_0^s=\frac{\tau_H\omega_s n_0}{4\pi\nu_z}\Big(G_0^s(z)-\nu_z \mathrm{A_1}q_0^s\Big).
	\end{equation}
	
	We take $L_{0}^{s} = L_{0}^{s^c} + L_{0}^{s^d}$. Then, the collimated component $L_{0}^{s^c}$ satisfies
	\beq
	\frac{\partial L_0^{s^c}}{\partial z}+\left(\frac{\tau_H n_s}{\nu_z}\right)L_0^{s^c} =0, \label{32}                
	\eeq   
	where the top boundary condition
	\beq\
	L_{0}^{s^c}(z=1,\theta) =  \delta(\nu_{z}-\nu_{z_0})\delta(\phi-\phi_0). 
	\eeq
	
	Now solving Equations (\ref{32}), we get
	$$
	L_{0}^{s^c}=\mathrm{I_t}\exp\left(\int_z^1\tau_H
		n_s(z')\,\ud z'\right)\delta(\nu_{z}-\nu_{z_0})\delta(\phi-\phi_0).
	$$
	
	Next the diffuse component $L_{0}^{s^d}$ satisfies
	\beq
	\frac{\p L_{0}^{s^d}}{\p z} + \frac{\tau_H n_s L_{0}^{s^d}}{\nu_{z}} =\frac{\omega_s\tau_H
		n_s}{4\pi\nu_{z}}{\Big(}G_{0}^{s}(z)-\mathrm{A_1} q_{0}^{s} \,\nu_{z}{\Big)},                
	\eeq      
	where the boundary conditions
\begin{numparts}
	\begin{eqnarray}
	L_{0}^{s^d}(z=1,\theta)=0,\quad \theta\in[\pi/2,\pi],\\
	L_{0}^{s^d}(z=0,\theta)=0,\quad  \theta\in[0,\pi/2].
		\end{eqnarray}
\end{numparts}
	
	Now we consider 
	$$
	\varpi = \int_z^1 \tau_H n_s(z')\,\ud z',
	$$ 
	thus $G_{0}^{s}$ and ${\bmi q}_{0}^{s}$ depend on $\varpi$ only.  
	
	Next, the total intensity at the equilibrium state is determined by
	\begin{equation}\label{36}
	G_{0}^{s}=G_{0}^{s^c}+G_{0}^{s^d},
	\end{equation}
	where
	$$
	G_{0}^{s^c}=\int_0^{4\pi}L_{0}^{s^c}(z,\theta)\,\ud\Omega=\mathrm{I_t}\exp\left(-\tau_H\int_z^1
	n_s(z')\,\ud z'\right),$$ and
	$$
	G_{0}^{s^d} =\int_0^{4\pi}L_{0}^{s^d}(z,\theta)\,\ud\Omega.  
	$$
	
	Also, the intensity flux at the basic state is written as 
	\beq
	{\bmi q}_{0}^{s} =  {\bmi q}_{0}^{s^{c}}+{\bmi q}_{0}^{s^{d}}, \label{37}
	\eeq
	where
	$$ 
	{\bmi q}_{0}^{s^{c}} = \int_0^{4\pi}L_{0}^{s^c}\,{\bmi s}\,\ud\Omega=-\mathrm{I_t}\exp
	\left(-\tau_H\int_z^1 n_s(z')\,\ud z'\right)\hat{\bmi z},$$ and
	$$
	{\bmi q}_{0}^{s^{d}} = \int_0^{4\pi}L_{0}^{s^d}(z,\theta)\,{\bmi s}\,\ud\Omega. 
	$$
	
	Equations~(\ref{36}) and (\ref{37}) are reformulated into coupled Fredholm integral equations~\cite{29modest2003} using the variable $\varpi$, i.e., 
	%\begin{widetext}
	\begin{eqnarray}
	\nonumber\fl G_{0}^{s}(\varpi) = \exp({-\varpi})+
	\frac{\omega_s}{2}\int_0^{\varpi}\Big(
	G_{0}^{s}(\varpi')\,E_1\left(|\varpi-\varpi'|\right) \\~~~~~~~~~~~~~~~~~~~~~~~~~~~~~~~~~~~~~~~
	+\mathrm{A_1} \,\textnormal{sgn}(\varpi-\varpi')q_{0}^{s}(\varpi')E_2\left(|\varpi-\varpi'|\right)\Big)\ud\varpi',
	\end{eqnarray}
	\begin{eqnarray}
	\nonumber\fl q_{0}^{s}(\varpi)=\exp({-\varpi })+\frac{\omega_s}{2}
	\int_{0}^{\varpi}\Big(\mathrm{A_1}\,q_{0}^{s}(\varpi')E_3\left(|\varpi-\varpi'|\right)\\~~~~~~~~~~~~~~~~~~~~~~~~~~~~~~~~~~~~~~~+\textnormal
	{sgn}(\varpi-\varpi')G_{0}^{s}(\varpi')E_2(|\varpi-\varpi'|)\Big)\ud\varpi', 
	\end{eqnarray}
	%\end{widetext}
	where $sgn$ represents the sign function, and $E_n(x)$ stands for the exponential integral of $n$ order~\cite{30chandrasekhar1960}. Also, these equations are tackled by employing the subtraction of singularity approach~\cite{31press1992,33crosbie1985}. 
	
	Therefore, in the equilibrium state, the mean orientation of microbes(algae here) is given by 
	$$
	\langle{\bmi p}_0^s\rangle=-M_0^s\frac{{\bmi q}_{0}^{s}}{q_{0}^{s}}=M_0^s\hat{\bmi z},
	\qquad\mbox{where}\quad
	M_0^s= M(G_{0}^{s}).
	$$ 
	
	The concentration at the equilibrium state satisfies 
	\beq
	\frac{\ud n_s(z)}{\ud z} = V_c\,M_0^s\,n_s,\label{40}
	\eeq
	which is supplemented by the cell conservation relation
	\beq
	\int_{0}^{1} n_s(z')\,\ud z' = 1.\label{41}
	\eeq 
	
	Equations~(\ref{40})--(\ref{41}) form a boundary value problem, which is solved using shooting method~\cite{31press1992}.  
	
	Also, at the equilibrium state the temperature profile satisfies
	\begin{equation}
	\frac{\ud^2 T_s(z)}{\ud z^2}=0,
	\end{equation}
    which is solved analytically using boundary conditions (\ref{24}) and (\ref{25}), thus we get	
	\begin{equation}
	T_s(z)=1-z.
	\end{equation}

	We utilize the subsequent collection of parameters as outlined by Panda \textit{et al.}~\cite{16panda2022} and Panda~\cite{8panda2020} in a proposed investigation: $S_c = 20,\mathrm{I_t}=1,\omega_s\in[0,1],\tau_{H}=0.5,1.0,$ $V_c =10,15,20$ and $\mathrm{A_1}\in[0,0.8]$ for finding the equilibrium solution.

	%%%%%%%%%%%%%%%%%%%%%%%%%%%%%%%%%%%%%%%%%%%%%%%%%%	
	\section{The perturbed system}\label{sec4}
	
	Now, the steady bioconvective system is perturbed by introducing small disturbances as
	%\begin{widetext}
	\begin{eqnarray*}
	{\bmi u}=(0,0,0)+\epsilon {\bmi u}_1+\mathcal{O}(\epsilon^2),\quad n=n_s(z)+\epsilon n_1+\mathcal{O}(\epsilon^2),\\\langle{\bmi
		p}\rangle=\langle{\bmi p}_{0}^{s}\rangle+ \epsilon \langle{\bmi p}_{0}^{1}\rangle +\mathcal{O}(\epsilon^2),\quad T=T_s+\epsilon T_1+\mathcal{O}(\epsilon^2), 
	\end{eqnarray*}
	%\end{widetext}
	where $0<\eps\ll 1.$ 
	
	Therefore, the system under linear perturbation is expressed as
	\beq
	\bnabla\bcdot{\bmi u}_1=0,\label{44}
	\eeq
	\beq
	P_r^{-1} \frac{\p {\bmi u}_1}{\p t} = 
	\nabla^2{\bmi u}_1-\bnabla P_e - R n_1 \hat{\bmi z}+R_TT_1\hat{\boldsymbol{z}},\label{45}
	\eeq
	\beq
	\frac{\p n_1}{\p t} + w_1 \frac{\ud n_s}{\ud z} +\frac{1}{Le}V_c\,\bnabla \bcdot(\langle{\bmi
		p}_{0}^{s}\rangle n_1 + \langle{\bmi p}_{0}^{1}\rangle n_s)=\frac{1}{Le}\nabla^2 n_1. 
	\eeq
	\beq
	\frac{\partial{T_1}}{\partial t}-w_1\frac{\ud T_s}{\ud z}=\boldsymbol{\nabla}^2T_1.\label{47}
	\eeq
	
	Let $L_{0}^{1^c}$ and $L_{0}^{1^d}$ be the perturbed collimated and diffuse intensities, i.e., $L_{0}=(L_{0}^{s^c}+L_{0}^{s^d})+\eps (L_{0}^{1^c}+L_{0}^{1^d})+\mathcal{O}(\epsilon^2)$. If the total intensity 
	$G=G_{0}^{s}+\eps G_{0}^{1}+\mathcal{O}(\epsilon^2)$, then
	\begin{equation}\label{48}
	G_{0}^{1}=G_{0}^{1^c}+G_{0}^{1^d},
	\end{equation}
	where
	$$
	G_{0}^{1^c} =\mathrm{I_t} \frac{\tau_H}{\cos{\theta_{0}}} \left(\int_1^zn_1\,\ud z' \right)\exp\left(-\frac{\tau_H}{\cos{\theta_{0}}}\int_z^1
	n_s(z')\,\ud z'\right),$$ and
	$$
	G_{0}^{1^d} =\int_0^{4\pi}L_{0}^{1^d}({\bmi x},{\bmi s})\,\ud\Omega.  
	$$
	
	Also, the radiative heat flux is ${\bmi q}={\bmi q}_{0}^{s}+\eps {\bmi q}_{0}^{1}+\mathcal{O}(\epsilon^2)$, then
	\begin{equation}\label{49}
	{\bmi q}_{0}^{1}={\bmi q}_{0}^{1^c}+{\bmi q}_{0}^{1^d},
	\end{equation}
	where
	%\begin{widetext}
	$$
	{\bmi q}_{0}^{1^c} = -\mathrm{I_t}\frac{\tau_H}{\cos{\theta_{0}}} \left(\int_1^zn_1\,\ud z' \right)\exp\left(-\frac{\tau_H}{\cos{\theta_{0}}}\int_z^1
	n_s(z')\,\ud z'\right)\cos{\theta_{0}}\hat{\bmi z},$$ 
	and
	$$
	{\bmi q}_{0}^{1^d} =\int_0^{4\pi}L_{0}^{1^d}({\bmi x},{\bmi s})\bmi s\,\ud\Omega.  
	$$
	%\end{widetext}
	
	By simplifying the expression
	$$
	-M(G_{0}^{s}+\eps G_{0}^{1}+\mathcal{O}(\epsilon^2))\frac{{\bmi q}_{0}^{s} + \eps {\bmi q}_{0}^{1}+\mathcal{O}(\epsilon^2)}{|{\bmi q}_{0}^{s} + \eps {\bmi q}_{0}^{1}+\mathcal{O}(\epsilon^2)|}-M_0^s\hat{\bmi z},
	$$
	the perturbed swimming direction is determined as
	\beq
	\langle{\bmi p}_{0}^{1}\rangle=G_{0}^{1}\frac{\ud M_0^s}{\ud G} \hat{\bmi z}-M_0^s\frac{{\bmi q}_{0}^{1^H}}{q_{0}^{s}}, 
	\eeq
	where ${\bmi q}_{0}^{1^H}=({\bmi q}_{0}^{1^x},{\bmi q}_{0}^{1^y})$. 
	
	Now, we eliminate the pressure term from Equation (\ref{45}) and horizontal components of $\boldsymbol{u}_1$ by taking double curl and retaining vertical component only. Hence, Equations~(\ref{44})--(\ref{47}) are reduced to set of equations for $w_1,n_1$ and $T_1$. These quantities can then be decom-
	posed into normal modes such that
	\begin{equation}
	\fl w_1=W(z)f(x,y) \exp({\gamma t}), \quad n_1=\Theta(z)f(x,y) \exp({\gamma t}), \quad T_1=T(z)f(x,y) \exp({\gamma t})
	\end{equation}
	where $f(x,y)=\exp[{i(lx+my)}]$.
	
	The perturbed diffuse intensity $L_{0}^{1^d}$ satisfies
	%\begin{widetext}
	\begin{eqnarray}
	 \nonumber\fl\xi_{x} \frac{\p L_{0}^{1^d}}{\p x} + \eta_{y} \frac{\p L_{0}^{1^d}}{\p y} + \nu_{z} \frac{\p L_{0}^{1^d}}{\p
		z} + \tau_H n_s L_{0}^{1^d}=\frac{\omega_s\tau_H}{4\pi}\Big(n_s G_{0}^{1} + G_{0}^{s}
	n_1+\mathrm{A_1}\nu_{z}\left(n_s{\bmi q}_{0}^{1}\bcdot{\hat{\bmi z}}- q_{0}^{s} n_1\right)\Big)\\~~~~~~~~~~~~~~~~~~~~~~~~~~~~~~~~~~~~~~~~~~~~~~~~~~~~~~~~~~~~~~-\tau_H L_{0}^{s} n_1,\label{52}
	\end{eqnarray}  
	%\end{widetext}
	subject to the boundary conditions	
	\begin{numparts}
		\begin{eqnarray}
		L_{0}^{1^d}(x,y,z=1,\xi_{x},\eta_{y},\nu_{z})=0,\,\theta\in[\pi/2,\pi],\, \phi\in[0,2\pi],\\
		L_{0}^{1^d}(x,y,z=0,\xi_{x},\eta_{y},\nu_{z})=0,\, \theta\in[0,\pi/2],\, \phi\in[0,2\pi].
		\end{eqnarray}
	\end{numparts}
	
	$L_{0}^{1^d}$ can be expressed as (see Equations~(\ref{52}))
	\beq
	L_{0}^{1^d}=\mathcal{\Psi}(z,\xi_{x},\eta_{y},\nu_{z}) f(x,y) \exp({\gamma t}). \label{eq38}   
	\eeq
	
	Also, Equations~(\ref{48}) yields
	%\begin{widetext}
	\begin{eqnarray}
	\nonumber\fl G_{0}^{1^c} =\mathrm{I_t}\left\{\frac{\tau_H}{\cos{\theta_{0}}} \left(\int_1^z\Theta(z')\,\ud z' \right)\exp\left(-\frac{\tau_H}{\cos{\theta_{0}}}\int_z^1
	n_s(z')\,\ud z'\right)\right\} f(x,y) \exp({\gamma t})\\~~~~~~~~~~~~~~~~~~~~~~~~~~~~~~~~~~~~~~~~~~~~~~~~~~~~~~~~~~~~~~~~~~~~~=\mathcal{G}_0^{1^c}(z)f(x,y) \exp({\gamma t}),\qquad
	\end{eqnarray}
	\begin{equation}
	\fl G_{0}^{1^d}=\mathcal{G}_0^{1^d}(z)f(x,y) \exp(\gamma t)=\bigg(\int_0^{4\pi}\mathcal{\Psi}(z,\xi_{x},\eta_{y},\nu_{z})\bigg) f(x,y) \exp({\gamma t}),
	\end{equation}
	%\end{widetext}
	where $
	\mathcal{G}_{0}^{1}(z)=\mathcal{G}_{0}^{1^c}(z)+\mathcal{G}_{0}^{1^d}(z)$.
	
	From Equation (\ref{49}), we get perturbed radiative heat flux
	$$
	{\bmi q}_{0}^{1}=\left[q_{0}^{1^x},q_{0}^{1^y},q_{0}^{1^z}\right]=\left[Q_x(z),Q_y(z),Q_z(z)\right] f(x,y) \exp({\gamma t}), 
	$$
	where 
	\beq
	\left(Q_x(z),Q_y(z)\right) =
	\int_0^{4\pi}\big(\xi_{x},\eta_{y}\big)\mathcal{\Psi}(z,\xi_{x},\eta_{y},\nu_{z})\,\,\ud\Omega, 
	\eeq
	and 
	$$
	Q_z(z)=-\mathcal{G}_{0}^{1^c}+\int_0^{4\pi}\nu_{z}\mathcal{\Psi}(z,\xi_{x},\eta_{y},\nu_{z})\,\,\ud\Omega.
	$$
	
	Now from equation (\ref{52}), $\mathcal{\Psi}$ satisfies
	%\begin{widetext}
	\begin{equation}
	\fl\frac{\p \mathcal{\Psi}}{\p z} + \frac{i(l\xi_{x}+m\eta_{y})+\tau_H n_s}{\nu_{z}}\mathcal{\Psi}=
	\frac{\omega_s \tau_H}{4\pi\nu_{z}}\bigg(n_s \mathcal{G}_{0}^{1} + G_{0}^{s}\Theta+\mathrm{A_1}\nu_{z}\left(n_s Q_z-
	q_{0}^{s}\Theta\right)\bigg)-\frac{\tau_H}{\nu_z}L_{0}^{s}\Theta,
	\end{equation}
	%\end{widetext}
	with the boundary conditions
	\begin{numparts}
		\begin{eqnarray}
		\mathcal{\Psi}\,(z=1,\xi_{x},\eta_{y},\nu_{z})=0,\quad \theta\in[\pi/2,\pi],\, \phi\in[0,2\pi],\\
	\mathcal{\Psi}\,(z=0,\xi_{x},\eta_{y},\nu_{z})=0,\quad \theta\in[0,\pi/2],\, \phi\in[0,2\pi].
		\end{eqnarray}
		\end{numparts}
	
	The perturbed governing equations become
	%\begin{widetext}
	\begin{equation}\label{60}
	\fl\left(\gamma P_r^{-1}+k^2-\frac{\ud^2}{\ud z^2}\right)\left(\frac{\ud^2}{\ud
		z^2}-k^2\right)W=R k^2 \Theta-R_Tk^2T,
	\end{equation}
	\begin{equation}
	\fl\left(\gamma Le+k^2-\frac{\ud^2}{\ud z^2}\right)\Theta + V_c \frac{\ud}{\ud z}\left(M_0^s\Theta+n_s\frac{\ud M_0^s}{\ud G_{0}^{s}}\mathcal{G}_{0}^{1}\right)-i\frac{V_c n_s M_0^s}{q_{0}^{s}}(lQ_x+mQ_y) = -Le\frac{\ud n_s}{\ud z}W,
	\end{equation}
	\begin{equation}\label{62}
	\fl\left(\frac{\ud^2}{\ud z^2}-\gamma -k^2\right)T=\frac{\ud T_s}{\ud z}W,
	\end{equation}
	%\end{widetext}
	with the boundary conditions
	\begin{equation} 
	W=\frac{\ud\Theta}{\ud z}-V_c M_0^s \Theta - V_c n_s
	\frac{\ud M_0^s}{\ud G_s}
	\mathcal{G}_{0}^{1}=0, \,~~~ \mbox{at}\,~~~ z=0, 1.\label{63}
	\end{equation}
	For a rigid boundary
	\begin{equation} 
	\frac{\ud W}{\ud z}=0, \,~~~ \mbox{at}\,~~~ z=0, 1,
	\end{equation}
	for a stress free boundary
	\begin{equation} 
	\frac{\ud^2 W}{\ud z^2}=0, \,~~~ \mbox{at}\,~~~ z=1,
	\end{equation}
	and for temperature
	\begin{equation} 
	T=0, \,~~~ \mbox{at}\,~~~ z=0,1.
	\end{equation} 
	
	Here, the parameter $k=\sqrt{l^2+m^2}$ is a non-dimensional wavenumber. Equations~(\ref{60})--(\ref{62}) are formed an eigenvalue problem for the growthrate of the disturbance $\gamma$ as a function of $S_c$, $\mathrm{I_t},\tau_H$, $V_c$, $\omega_s,$, $k$,  $\mathrm{A_1}, R$, and $R_T$. It is noteworthy that if $\mathrm{Re}(\gamma)>0$, the system becomes unstable. 
	
	Now, we incorporate a new variable 
	\beq
	\widehat{\Theta}(z)=\int_1^z \Theta(z')\,\ud z'. 
	\eeq
	
	Consequently, the perturbed governing equations become (using $D=\ud/\ud z$) 
	%\begin{widetext}
	\begin{equation}\label{68}
	\left(D^2-k^2\right)\left(\gamma P_r^{-1}+k^2-D^2\right)W=R k^2
	D \widehat{\Theta}-R_Tk^2T,
	\end{equation}
	\begin{equation}
	D^3 \widehat{\Theta}-V_c M_0^s\,D^2\widehat{\Theta}-\big[\gamma Le+k^2+\widetilde{\Gamma}_2(z)\big]D \widehat{\Theta}-\widetilde{\Gamma}_1(z)\widehat{\Theta}-\widetilde{\Gamma}_0(z) =Le\, Dn_s W, 
	\end{equation} 
	\begin{equation}\label{70}
	\left(D^2-k^2-\gamma\right)T=\uD T_s\, W,
	\end{equation}
	%\end{widetext}
	where 	
	\begin{eqnarray}
	\widetilde{\Gamma}_0(z)&=&V_c\,\uD\left(n_s\frac{\ud M_0^s}{\ud G}\mathcal{G}_{0}^{1^d}\right)-i\frac{V_cn_sM_0^s}{q_{0}^{s}}(lQ_x+mQ_y),\\
	\widetilde{\Gamma}_1(z)&=& \frac{\tau_H}{\cos{\theta_{0}}} V_c \uD\left(n_sG_{0}^{s^c} \frac{\ud M_0^s}{\ud G}\right),  \\
	\widetilde{\Gamma}_2(z)&=& 2\frac{\tau_H}{\cos{\theta_{0}}} V_c n_s G_{0}^{s^c} \frac{\ud M_0^s}{\ud G}+V_c
	\frac{\ud M_0^s}{\ud G}\uD G_{0}^{s^d}.
	\end{eqnarray}  
	
	All boundary conditions remain the same except Equation (\ref{63}) which becomes
	\begin{equation} 
	W=\uD^2\widehat{\Theta}-V_c \,M_0^s\,\uD\widehat{\Theta} - V_c n_s
	\frac{\ud M_0^s}{\ud G_s}
	\mathcal{G}_{0}^{1}=0, \,~~~ \mbox{at}\,~~~ z=0, 1.
	\end{equation}
	and there is requirement of an additional boundary condition, which is given by
	\beq
	\widehat{\Theta}(z)=0,\qquad \mbox{at}\quad z=1.
	\eeq
	%%%%%%%%%%%%%%%%%%%%%%%%%%%%%%%%%%%%%%%%%%%%%%%%%%%%%%%%%%%%%%%%%%%%%%%%%%%%%%%%%%%%%%%%%%%%%%	
	\section{Solution procedure}\label{sec5}
	
	A finite-difference scheme of fourth-order accurate which iterates via the Newton–Raphson–Kantorovich (NRK) algorithm~\cite{18cash1980} is employed to solve Equations~(\ref{68})--(\ref{70}). This method helps to determine the specific parameters for which the basic state becomes unstable via diagramming the neutral (marginal) stability curves in the $(k,R)$-plane. The neutral stability curve is recognized as a graph of the points where $\textnormal{Re}(\gamma)=0$. If $\textnormal{Im}(\gamma)=0 \, (\textnormal{or} \neq 0)$  in such a graph, then the bioconvective instability is named as stationary (or oscillatory) one. The neutral curve $R^{(n)}(k),\, n\in\mathbb{N}$ possesses an infinite number of branches for any given set of parameters, and each branch is recognized as a solution to the linear stability problem. The most unstable mode of disturbance corresponds to the solution branch at which both $R=R_c$ and $k=k_c$. Here, the initial bioconvection pattern's wavelength is given by $\lambda_c=2\pi/k_c$. In addition, if a non-stationary branch of the neutral curve contains the most unstable mode of disturbance, then overstability solution occurs. In this instance, an oscillatory branch draws a plot for the set $\left\{k: k \le k_0\right\}$ at some point $k_0$. Also, the bioconvection solution is classified as mode $n$ when $n$ convective cells are arranged in a vertical stacking pattern, with one positioned above the other~\cite{8panda2020}. The proposed phototaxis model considers the governing parameters applicable to non-gyrotactic microbes similar to \textit{Chlamydomonas} [Refer Panda~\cite{8panda2020}].
	%%%%%%%%%%%%%%%%%%%%%%%%%%%%%%%%%%%%%%%%%%%%%%%%%%%%%%%%%%%%%%%%%%%%%%%%%%%%%%%
	\section{Numerical results}\label{sec6}
	
	A fixed parameter set is utilized to predict the most unstable mode of disturbance at bioconvective instability. These parameters are as follows: $P_r=5, \mathrm{I_t}=1.0$ along with $\tau_{H}=0.5,1.0,$ $\omega_s\in[0,1]$, and $V_c=10,15,20$. 
		\begin{figure}[!htbp]
		\begin{center}
			\includegraphics[width=7cm,height=6cm]{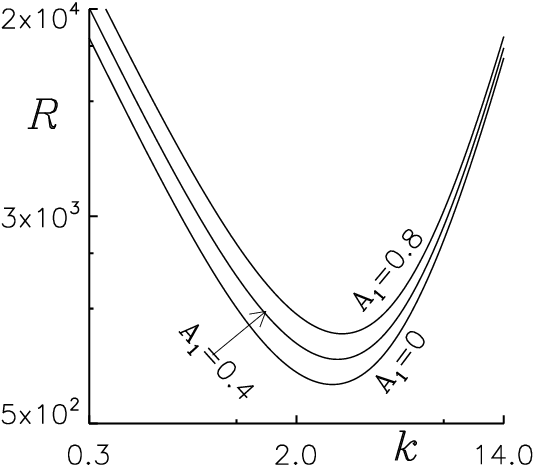}
		\end{center}
		\caption{Effects of forward scattering coefficient on the neutral curves. Here the parameters $V_c=15,G_c=1.0,\tau_H=0.5,\omega_s=0.4,Le=4$, and $R_T=100$ are kept fixed.}
		\label{fig3}
	\end{figure}
	%%%%%%%%%%%%%%%%%%%%%%%%%%%%%%%%%%%%%%%%%%%%%%%%%%%%%%%%%%%%%%%%%%%
		
	%\subsection{Weak-scattering suspension of microbes} \label{di}
	\subsection{$V_c=15$\newline} \label{}
	
	Figure \ref{fig3} shows the effects of forward scattering coefficient $\mathrm{A_1}$ on the neutral curves, where the parameters $V_c=15,G_c=1.0,\tau_H=0.5,\omega_s=0.4,Le=4$, and $R_T=100$ are kept fixed. Here, the critical bioconvective Rayleigh number increases as the forward scattering coefficient increases, which indicates that the suspension becomes more stable as the forward scattering coefficient increases.

\begin{figure}[!htbp]
	\begin{center}
		\includegraphics[scale=0.7]{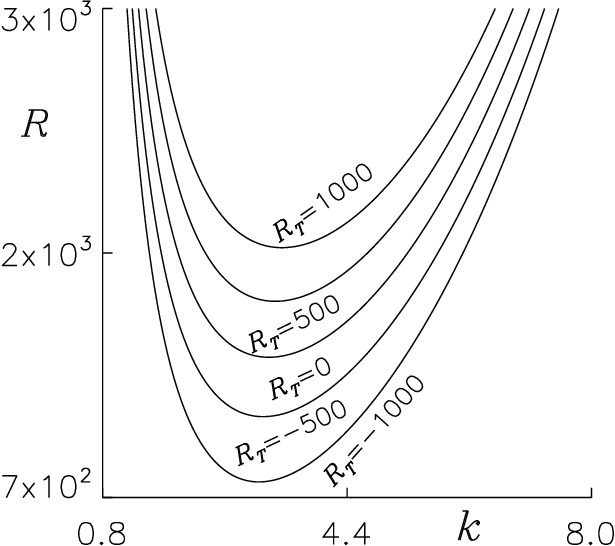}
	\end{center}
	\caption{Effects of thermal Rayleigh number at the neutral curves. Here the parameters $V_c=15,G_c=1.0,\tau_H=0.5,\omega_s=0.4,Le=4$, and $\mathrm{A_1}=0.4$ are kept fixed}
	\label{fig4}
\end{figure}

Figure \ref{fig4} demonstrates the effects of forward scattering coefficient $\mathrm{A_1}$ on the neutral curves, where the parameters $V_c=15,G_c=1.0,\tau_H=0.5,\omega_s=0.4,Le=4$, and $\mathrm{A_1}=0.4$ are kept fixed. Here, the critical bioconvective Rayleigh number increases as the thermal Rayleigh number increases, which indicates that the suspension becomes more stable as the Rayleigh number increases.

	%%%%%%%%%%%%%%%%%%%%%%%%%%%%%%%%%%%%%%%%%%%%%%%%%%%%%%%%%%%%%%%%%%%
	\section{Conclusions}\label{sec7}
	In this study, we introduce a new rational bioconvection model that consider the effects of both collimated irradiation and thermal gradient on bioconvection in a forward scattering suspension of microbes. The numerical results obtained via linear stability analysis of an algal suspension are summarized below.
	
	The total light intensity in a uniform weak-scattering algal suspension remains two-fold as the forward scattering coefficient escalates at a fixed angle of incidence. The profile of total intensity increases (or decreases) with respect to height in the lower (or upper) part of a weak-scattering suspension of microbes as the forward scattering intensifies. Similarly, the position of the peak of basic concentration profile transits towards the top in the upper part of a weak-scattering suspension as the forward scattering minimizes. Also, the highest value of the concentration profile intensifies (or degrades) in the lower (or upper) part of a weak-scattering suspension as the forward scattering intensifies. Forward scattering allows deep underwater light penetration and creates a limitlessness light environment within the cell suspension. For a purely scattering suspension, the bimodal steady cell concentration profile transforms into a unimodal one as the forward scattering coefficient increases.
		
	Now, we discuss about the predicted numeral results via linear stability analysis. The critical bioconvective Rayleigh number increases with an augmentation in a forward scattering coefficient when the thermal Rayleigh number is fixed. Similarly, the critical bioconvective Rayleigh number increases with an augmentation in a thermal Rayleigh number when the forward scattering coefficient is fixed. Thus, the suspension becomes more stable as the forward scattering coefficient and thermal Rayleigh number increase.
	%%%%%%%%%%%%%%%%%%%%%%%%%%%%%%%%%%%%%%%%%%%%%%%%%%%
%	\section*{Acknowledgements}
%	The second author expresses gratitude to the DST (SERB), Government of India, a statutory body, for the financial assistance provided by the Core Research Grant (Grant No. CRG/2018/000105).
	%%	
	\section*{Author decelerations}
	\section*{Conflict of interest}
	The author discloses no conflicts of interest.
	\section*{Data availability}
	The article contains all the data that supports the study's results.
	\section*{References}
	\bibliographystyle{iopart-num}
	\bibliography{IOP_ANISO_THERMAL_BIOCONVECTION}
\end{document}